\documentclass[prd, twocolumn, nofootinbib, floatfix]{revtex4-1}

\usepackage{amsmath}
\usepackage{graphicx}
\usepackage{dcolumn}
\usepackage{bm}
\usepackage{enumerate} 
\usepackage{epsfig}
\usepackage{amssymb,latexsym,mathrsfs}
\usepackage{graphicx}
\usepackage{color}
\usepackage{hyperref}

\hypersetup{
    colorlinks=true,
    linkcolor=red,
    citecolor=blue,
} 

\newcommand{\be}{\begin{equation}}
\newcommand{\ee}{\end{equation}}
\newcommand{\bes}{\begin{equation*}}
\newcommand{\ees}{\end{equation*}}
\newcommand{\beq}{\begin{equation}}
\newcommand{\eeq}{\end{equation}}
\newcommand{\ra}[1]{\renewcommand{\arraystretch}{#1}}
\newcommand{\bs}{\begin{split}} 
\newcommand{\bea}{\begin{eqnarray}}
\newcommand{\eea}{\end{eqnarray}}
\newcommand{\beqa}{\begin{eqnarray}}
\newcommand{\eeqa}{\end{eqnarray}}

\newcommand{\kmax}{k_{\rm max}}
\newcommand{\hmpc}{\,h\,{\rm Mpc}^{-1}}

\begin{document}

\title{Cosmological Leverage from the Matter Power Spectrum 
in the Presence of Baryon and Nonlinear Effects} 
\author{Jannis Bielefeld$^1$, Dragan Huterer$^2$, Eric V.\ Linder$^3$} 
\affiliation{$^1$Department of Physics \& Astronomy, Dartmouth College, 
Hanover, NH 03755 USA\\ 
$^2$Department of Physics, University of Michigan, 450 Church St, Ann Arbor, MI 48109-1040\\ 
$^3$Berkeley Center for Cosmological Physics \& Berkeley Lab, 
University of California, Berkeley, CA 94720, USA} 

\begin{abstract}
We investigate how the use of higher
wavenumbers (smaller scales) in the galaxy clustering power spectrum
influences cosmological constraints. We take into account uncertainties from
nonlinear density fluctuations, (scale dependent) galaxy bias, and baryonic
effects.  Allowing for substantially model independent 
uncertainties through separate fit parameters in each wavenumber bin that
also allow for the redshift evolution, we quantify strong gains in dark 
energy and neutrino mass leverage with increasing maximum wavenumber, despite
marginalizing over numerous (up to 125) extra fit parameters.  The leverage is
due to not only an increased number of modes but, more significantly, breaking
of degeneracies beyond the linear regime.
\end{abstract}

\date{\today} 

\maketitle

\section{Introduction} 

The statistical pattern of large scale structure in the universe contains 
a wealth of information on the cosmological parameters, including the nature 
of dark energy and the sum of neutrino masses. While the linear density 
perturbation power spectrum of 
dark matter can be related to the cosmological model in a straightforward 
manner, the observational data involves several complicating effects. We 
would like to use not only fully linear modes but the more numerous higher 
wavenumber modes where nonlinear effects appear; indeed the nonlinear regime 
contains not just more modes but distinct cosmological leverage. 

On these smaller scales, our understanding of the cosmological dependence 
is imperfect, while the statistical precision of large volume surveys can 
reach the subpercent level. Moreover, since we observe the light from 
galaxies, the mapping from dark matter predictions to data involves the 
galaxy bias factor, expected to be scale dependent beyond the linear regime. 
Finally, since galaxies contain dissipative baryons, various dynamical 
and feedback mechanisms not present for pure dark matter will alter the 
power spectrum. 

These nonlinearity, bias, and baryon effects can be addressed in a number of
ways, such as perturbation theory, the halo model formalism, and advanced
N-body and hydrodynamic computational simulations, with varying levels of
success.  
Considerable literature exists on these issues; for a selection see 
\cite{White:2004kv,Zhan:2004wq,Huterer:2004tr,Huterer:2005un,
Hagan:2005nb,Jing:2005gm,Rudd:2007zx,Zentner:2007bn,Hearin:2011bp,Zentner:2012mv}. 
For some recent work, especially regarding baryonic effects on the
weak lensing shear power spectrum, see
\cite{14056205,Eifler:2014iva,14074301,14106826}. 
The further we extend to higher wavenumbers, the less certain we are of having
captured all the necessary physics inputs, especially for the range of
cosmologies to be examined. One approach is simply to cut out the scales
beyond the (quasi)linear regime, using only wavenumbers up to some low
$\kmax$. This severely restricts the information used to a small fraction of
the data provided by the survey. An alternate approach is to focus on the
cosmological information and marginalize over the uncertain effects. This uses
more of the data, but the key danger here is assuming an improper functional
form for the unknown influences and so causing a systematic bias in the
cosmological conclusions.

We follow the marginalization approach but in a substantially model 
independent way, allowing the nonlinearity, galaxy bias, and baryon effects 
to float freely in bins of wavenumber without imposing a functional form 
for their scale dependence. This effectively removes 
the danger of distorting the cosmology results. 
The question then is whether the degradation in cosmological 
constraints due to the additional bin fit parameters outweighs the gain 
from including the further data. We focus on the real space matter power 
spectrum, for clarity in assessing the cosmological information content as 
a function of $\kmax$ and because it is central to a variety of different 
cosmological probes; it is given by the substantially transverse modes of 
a spectroscopic galaxy redshift survey, and enters in the angular galaxy 
power spectrum of 
photometric surveys and in the weak lensing shear power spectrum. 

In Sec.~\ref{sec:method} we describe our method of accounting for the scale 
and redshift dependence of the uncertain physics beyond the linear regime. 
We lay out the Fisher analysis approach and galaxy redshift survey 
characteristics in Sec.~\ref{sec:fisher}, then examine the behavior of 
the derived cosmological constraints as a function of $\kmax$ in 
Sec.~\ref{sec:results}. To test the robustness of the model independence, 
in Sec.~\ref{sec:fids} we consider alternative fiducials for the scale and 
redshift dependence. Appendix~\ref{sec:apx} further tests the approach by 
varying the binning properties. We discuss and summarize the results in 
Sec.~\ref{sec:concl}.

\section{Power Spectrum Effects} 
\label{sec:method} 

In the linear density perturbation regime, the real space dark matter power 
spectrum is readily given by Boltzmann codes such as CAMB \cite{Lewis:1999bs} 
or CLASS \cite{Blas:2011rf}. This can be extended beyond the linear regime 
through simulations or emulators built on simulations, e.g.\ 
\cite{cosmicemu,emu13}, 
or through nonlinear mapping of the linear power spectrum in algorithms 
such as Halofit \cite{Smith:2002dz} and its variants. Considerable work has 
recently gone into substantially extending perturbation theory and general 
wavenumber expansions to higher Fourier wavenumbers $k$, such as through 
the effective theory of large scale structure (see \cite{eftlss} and 
references therein) or the halo model (e.g.\ \cite{Mohammed:2014lja}). 
The status of these in accounting accurately for 
galaxy bias and baryon effects, over a range of cosmologies, is not yet 
clear though interesting progress is being made. 

Here we consider a phenomenological approach that does not rely on 
understanding fully the cosmological dependence of internal halo distributions 
or baryonic feedback. We write the power spectrum as 
\be 
P_X(k,z)=b_X^2(k,z)\,P_{\rm model}(k,z)\,M_{\rm baryon}(k,z) \ ,
	\label{eqn:model} 
\ee 
where $P_{\rm model}$ is some model for the (nonlinear) power spectrum 
whose cosmological 
dependence is well defined. This is multiplied by a factor $b_X^2$ describing 
the possibly scale dependent galaxy bias for some galaxy population $X$, and 
another function $M_{\rm baryon}$ dealing with baryonic effects. The 
separability of the factors is not essential, only for illustrative purposes. 

Uncertainties in the bias factor will be degenerate with those in the baryonic 
factor (unless specific functional forms are assumed), so we can absorb 
these both into the same factor, writing 
\be 
	P_X(k,z)=b_{X,{\rm fid}}^2(z)\,P_{\rm model}(k,z)\,M(k,z) \ . 
	\label{eq:p3} 
\ee 
Thus the uncertainties due to galaxy bias, nonlinearities, and baryonic 
effects are represented by the function $M(k,z)$. 
For compactness, we call $M$ the BNB factor, referring to 
all three sources of uncertainty. 
We will then marginalize 
over this and study the effect on the cosmological information extracted 
from the data. 

Our best guess, baseline dark matter power spectrum is $P_{\rm model}$ and 
this includes all the cosmological parameter dependence we will use. While 
$M$ may have further cosmological dependence, this will be lost in the 
marginalization, reducing the statistical leverage but guarding against 
systematic bias. We adopt for $P_{\rm model}$ the revised Halofit form of 
\cite{Takahashi:2012em}, updating the original \cite{Smith:2002dz}. 
Note that in the CAMB and CLASS versions from March 2014 and later this also 
includes the neutrino mass effects from \cite{Bird:2011rb}. 

To keep explicit 
the galaxy population dependence we retain a fiducial galaxy bias factor 
\be 
b_{X,{\rm fid}}(z)=b^0_{X,{\rm fid}}\,\frac{D_{\rm fid}(z=0)}{D_{\rm fid}(z)} 
\ , \label{eq:bias} 
\ee 
where $D_{\rm fid}$ is the growth factor for some fiducial cosmology. 
All deviations in the galaxy bias from this form, including scale dependence, 
enter in $M$ (as do deviations from the Halofit prescription for nonlinearity, 
and baryonic effects, i.e.\ the BNB effects). 

To keep $M(k,z)$ as general as possible to account for these uncertainties, 
we allow it to float freely in bins of wavenumber $k$, so that the data 
determines its form and amplitude. This approach worked well in exploration 
of the anisotropic, redshift space power spectrum uncertainties (without 
baryon or scale dependent bias effects) and its cosmological leverage in 
\cite{Linder:2012xv}. The redshift dependence of $M$ should be 
reasonably smooth as galaxy bias, excess nonlinearity, and baryonic effects 
develop on a roughly Hubble time scale. 

Our prescription is therefore 
\be 
M(k,z)=(1+c_{1,k}z+c_{2,k}z^2)\,B_k \ , 
\label{eqn:Mkz}
\ee 
where $B_k$ is an orthogonal bin basis with width $\Delta k=0.025\hmpc$, 
and $c_{i,k}$ are free parameters. 
(See Appendix~\ref{sec:apx} for tests of varying the bin width, and 
Sec.~\ref{sec:fidz} for extending the redshift dependence.) 
This gives 3 free parameters per bin; as we extend the maximum wavenumber 
$\kmax$ used from the data, we include more modes but also add more fit 
parameters 
to account for the further uncertainty. For example, assuming the uncertainty 
starts beyond the linear regime $k_{\rm low}=0.05\hmpc$, then including 
modes out to $\kmax=0.5\hmpc$ would add 54 fit parameters (plus $b^0_X$ for 
each galaxy population, plus cosmological parameters).

\section{Parameters and Information} \label{sec:fisher} 

\subsection{Clustering Information}

We perform a Fisher matrix analysis to compute the uncertainties and 
covariances of the various cosmological and astrophysical parameters of our 
model 
Eq.~(\ref{eq:p3}). This allows us to project the expected constraints from 
upcoming survey data, including as a function of $\kmax$. 

The full set of parameters $\{\theta_i\}$ includes cosmological parameters, 
fiducial galaxy biases, and the BNB parameters that enter through 
Eq.~(\ref{eqn:Mkz}). Each $k$ bin introduces three parameters 
\begin{equation}
{\vec\theta}_{\rm BNB}=\{B_k,c_{1,k}, c_{2,k}\} \ . 
\end{equation}
In our analysis we will derive constraints for 
survey samples of emission line galaxies (ELG) and luminous red galaxies 
(LRG), which add an extra two fiducial bias parameters 
\begin{equation}
{\vec\theta}_{\rm bias}=\{b^0_{ELG}, b^0_{LRG}\} 
\end{equation}
through Eq.~(\ref{eq:bias}) [recall that galaxy bias scale and redshift
  dependence beyond the fiducial is absorbed into $M(k,z)$]. Finally, the cosmological
model itself involves nine parameters:
\begin{equation}
{\vec\theta}_{\rm cosmo}=
\{\Omega_b h^2, \Omega_{\rm CDM}h^2, \Omega_\nu h^2, \Omega_K, h, w_0, w_a, A_s, n_s \}, 
\end{equation}
the physical baryon, cold dark matter, and neutrino energy densities, the
spatial curvature effective density, the reduced Hubble constant, the dark
energy equation of state parameters, and the amplitude and tilt of the
primordial density power spectrum. We call these three sets the BNB, bias, and
cosmological parameters; the full parameter space is the union of the three.
The fiducial values for these parameters are summarized in
Table~\ref{tab:fid}.

\begin{table*}
\ra{1.3}
	  \begin{tabular*}{0.75\textwidth}{@{\extracolsep{\fill} }c c c c c c c c c c c c c c}
	      $\Omega_b h^2$ & $\Omega_{\rm CDM}h^2$  & $\Omega_\nu h^2$ & $\Omega_K$ & $h$ & $w_0 $ & $w_a$ & $10^9 A_s$ & $n_s$ & $b^0_{\rm ELG} $  & $b^0_{\rm LRG}$ & $B_k$ & $c_{1,k}$ & $c_{2,k}$ \\ \hline
	    0.0226 & 0.112 & 0.00064 & 0 & 0.7 & -1 & 0 & 2.19 & 0.96 & 0.8 & 1.6 & 1 & 0 &0 \\
	  \end{tabular*}
	  \caption{Fiducial parameter values. The neutrino density 
corresponds to masses $\sum m_\nu=0.06\,eV$.}
	 \label{tab:fid}
\end{table*}

Galaxy clustering information in the form of the galaxy power spectrum 
contains cosmological information as prescribed in, e.g., 
\cite{Feldman:1993ky, Seo:2003pu, Stril:2009ey}. The error covariance 
matrix is assumed to be diagonal and only contains contributions from the 
sample variance and shot noise. The statistical error per Fourier mode is 
$\sigma_{P,{\rm mode}} = P + 1/n$ from these two effects where $P$ is the power 
spectrum and $n$ the shot noise. Upon division by the number of modes 
one obtains the error variance in the power spectrum
\be 
	\sigma_P^2=P^2\left(\frac{1+nP}{nP}\right)^2 \frac{8\pi^2}{V_{\rm shell}(z)\,k^2 dkd\mu} \ , 
	\label{eq:cov} 
\ee 
taking the $k$-modes as independent. 
This feeds into the Fisher matrix \cite{tegtayh} 
\be
F_{ij} = \sum_z \sum_k^{k_{\rm max}} \sum_{XY}\frac{\partial P_X(k,z)} 
{\partial \theta_i}\, \frac{1}{\sigma_P^2}\, \frac{\partial P_Y(k,z)}{\partial \theta_j} 
\label{eqn:fisher}
\ee 
where the $z$ and $k$ sums are over shells in the two variables, and the sum over $\mu$ is implicit.
Note that the factors $P$ from the error $\sigma_P$ combine with the 
derivatives to form logarithmic Fisher derivatives 
$\partial\ln P(k,z)/\partial\theta_i$. This is useful in numerically treating 
multiplicative factors; for example the numerous bin parameters for the 
BNB effects do not require additional calls 
to CAMB. 

The information is summed over $k$ modes out to some $\kmax$; one of our 
main aims is to investigate how the cosmological constraints coming from this 
added information -- but also with added free parameters in each $k$ bin -- 
behave 
as a function of $\kmax$. Note that the power spectrum can float freely in 
each $k$ bin (above $k_{\rm low}=0.05\,h$/Mpc), though with constrained 
redshift dependence given by Eq.~(\ref{eqn:Mkz}). In Sec.~\ref{sec:fids} 
we explore both enlarging this freedom and changing the fiducial model. 
While a change in the parameters $\{B_k,c_{1,k},c_{2,k}\}$ in one bin has no 
effect on the power spectrum in another bin (i.e.\ no model dependence is 
forced; the bins can float freely), the Fisher analysis will quantify the 
covariance between these variations given the data. All parameter 
uncertainties quoted have been marginalized over all other parameters.

\subsection{Survey Observables} \label{sec:obs}

The Fisher sum also extends over redshift shells $z$, 
with the survey shell volume and galaxy number densities $n(z)$ in each 
population $X$ or $Y$ changing with redshift. We consider a next generation, 
``Stage IV'' galaxy redshift survey of the quality planned from DESI 
\cite{desi} or Euclid \cite{euclid}, 
specifically adopting the $n(z)$ from \cite{Linder:2012xv}, covering 
$z=0.1$--1.8 over 14000 deg$^2$. 

One caveat is that our focus is a theoretical study of the information 
content in the real space galaxy power spectrum. While this is a key 
ingredient 
in many observations -- the redshift space power spectrum, the angular 
power spectrum, the weak lensing power spectrum, etc.\ -- it is not directly 
observable. One should therefore view the results as a theoretical analysis 
of the innate information. Alternately, one might expect that the 
{\rm relative\/} behavior of the cosmological constraints with $\kmax$, if 
not their {\it absolute\/} values, still holds for, e.g., projection to 
an observable angular power spectrum. Another view is to say that 
redshift surveys do indeed measure the real space power spectrum for 
Fourier modes nearly transverse to the line of sight, and so one could 
include only $|\mu|<0.1$ [since corrections between real and redshift 
space go as $(k\mu)^2$ for small $k\mu$, this cutoff should give 
$\lesssim1\%$ accuracy] 
in the mode sum of Eq.~(\ref{eq:cov}); such an ansatz would reduce the 
information content used here uniformly by a factor 10, and all quoted 
cosmology constraints would increase by $\sqrt{10}\approx3.2$. 
In general, using a subset of the information in $P(k, \mu)$, or the
presence of other systematic errors not studied here (e.g.\ errors in
the measurements of shapes of weakly lensed galaxies), will  weaken the
overall cosmological constraints -- and therefore correspondingly weaken the
requirements on the selfcalibration of the BNB errors studied in this work.

\section{Results} \label{sec:results} 

As we include power spectrum information from higher $k$ bins, three 
effects enter: more modes are included, lowering the statistical uncertainty, 
more fit parameters are included (e.g.\ 54 more for $\kmax=0.5\,h$/Mpc), 
increasing the cosmological parameter uncertainty, and a longer lever arm 
on the Fisher derivatives is created, potentially breaking parameter 
degeneracies and decreasing the parameter uncertainty. To investigate which 
wins out, we compute the constraints for $k_{\rm max}=0.1$, 0.2, 0.3, 0.5, 
0.75, and $1\,h$/Mpc. 

First, note that in the linear regime, the cosmic growth of structure 
is scale independent and so late time cosmological parameters such as the 
dark energy parameters have Fisher derivatives $\partial\ln P/\partial 
\theta$ independent of $k$. This means that these $k$ modes cannot break 
degeneracies between such parameters (only redshift dependence can) and 
so marginalized uncertainties are quite high compared to unmarginalized 
ones. As we add higher $k$ information, however, beyond the linear regime, 
the derivatives gain different $k$-dependent shapes, 
breaking degeneracies and 
potentially allowing rapid improvement in parameter estimation. Of course 
if the BNB effects were incorrectly modeled, then the parameter estimation 
will be biased -- hence we employ the (substantially) model independent binned 
approach to avoid this, albeit at the price of adding more free parameters. 

\begin{figure}[htbp!]
\includegraphics[width=\columnwidth]{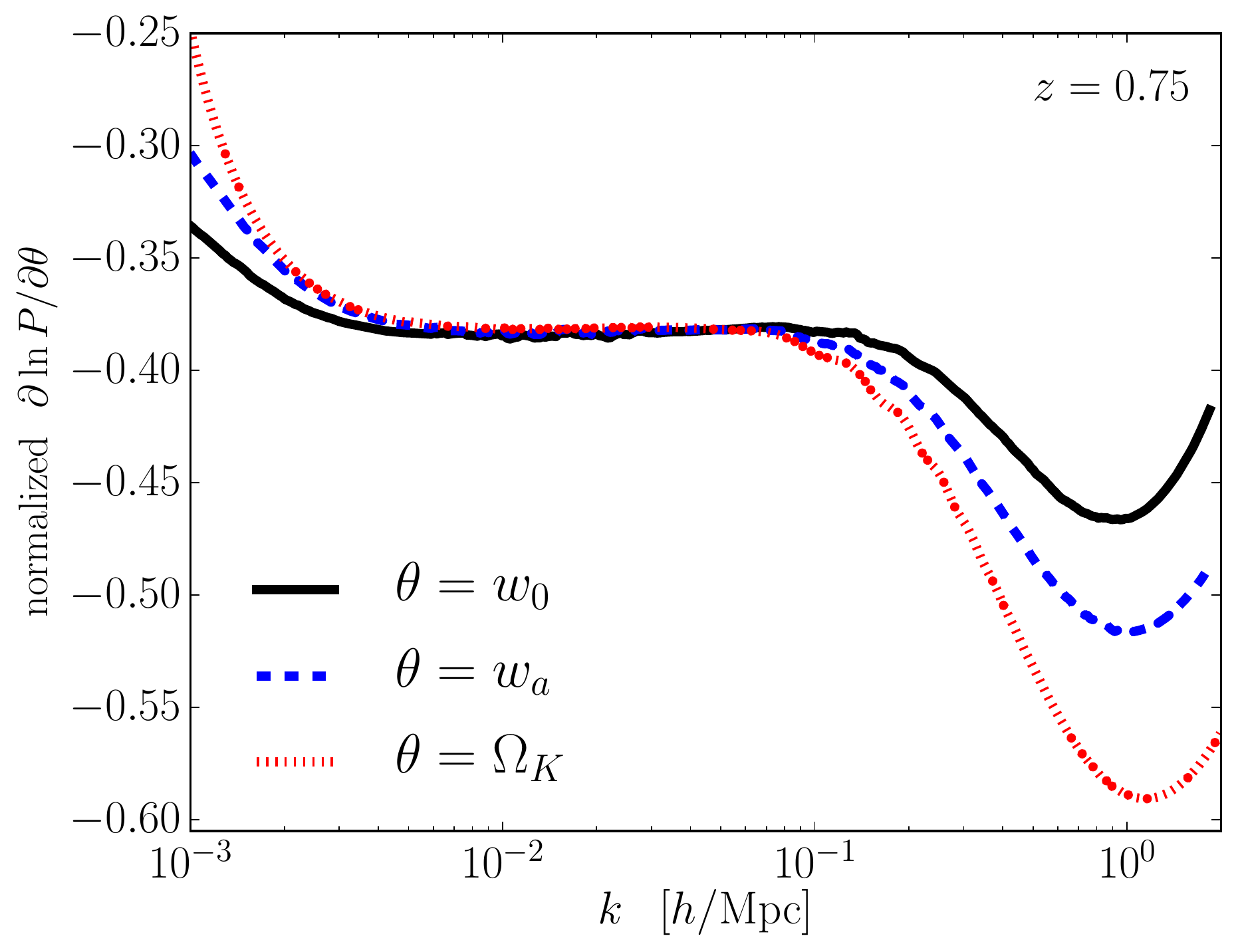}\\ 
\includegraphics[width=\columnwidth]{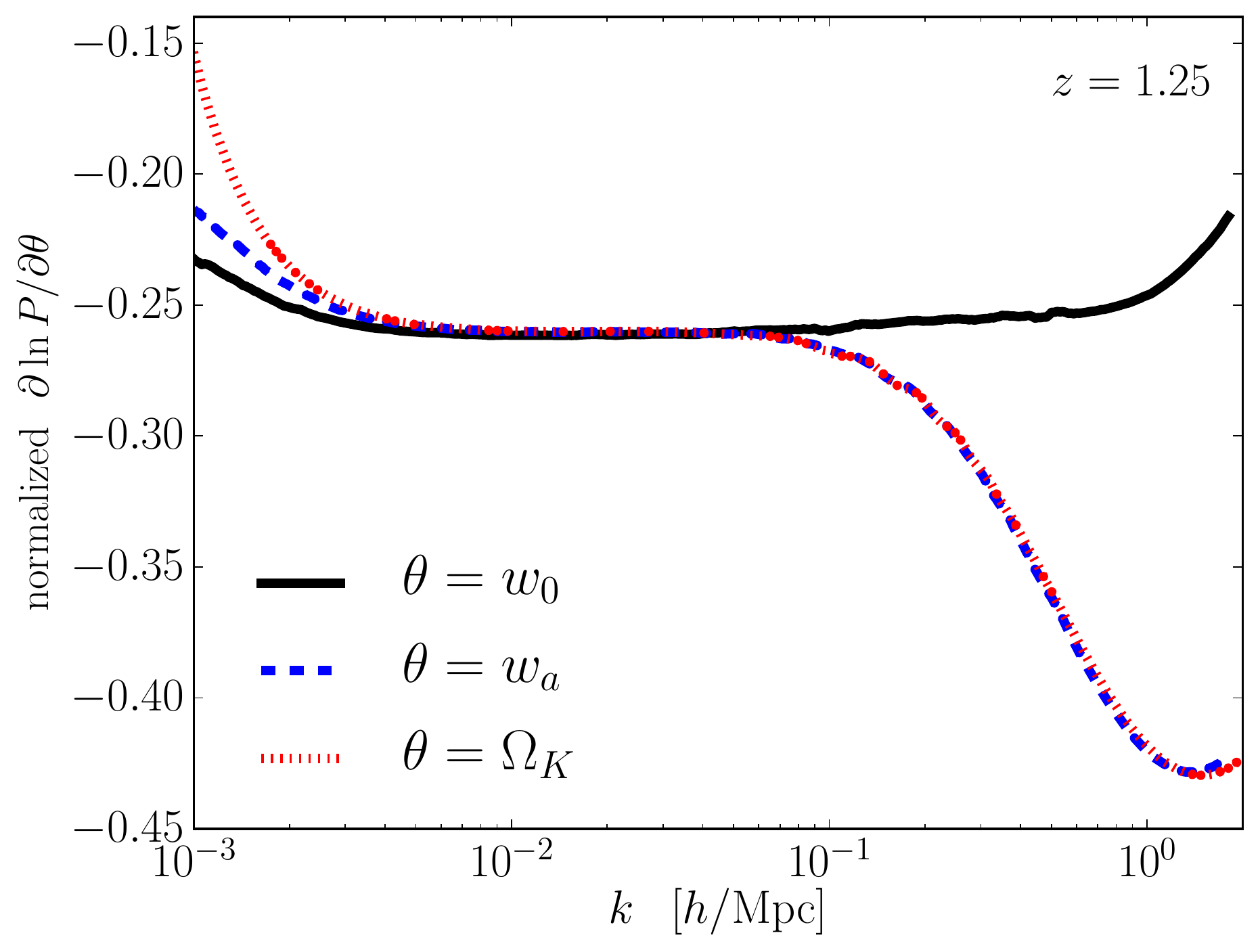} 
\caption{The  derivatives $\partial\ln P(k)/\partial\theta$ are 
plotted vs $k$ for $\theta=\{w_0,w_a,\Omega_K\}$, at $z=0.75$ (top) and 
$z=1.25$ (bottom). For clearer comparison 
of shapes, the $w_a$ and $\Omega_K$ curves are normalized to the value 
of the $w_0$ curve at $k=0.05\,h$/Mpc.
}
	\label{fig:degen} 
\end{figure}

Figure~\ref{fig:degen} plots the Fisher derivatives for several parameters 
as a function of $k$, normalized to $k=0.05\,h$/Mpc to highlight the 
shapes. We see that beyond $k\approx0.1\,h$/Mpc the curves for the 
late time parameters such as the dark energy equation of state and curvature 
begin to diverge, lowering their covariance with each other. Interestingly, 
at redshifts $z\approx1.2$--1.6 substantial covariance extends to higher 
$k$ for $w_a$ and $\Omega_K$, suggesting that high-redshift galaxy 
clustering surveys would benefit from combination with lower redshift 
surveys, or that a survey should span both low and high redshift for best 
results.

\begin{figure*}[htbp!]
	\includegraphics[width=\columnwidth]{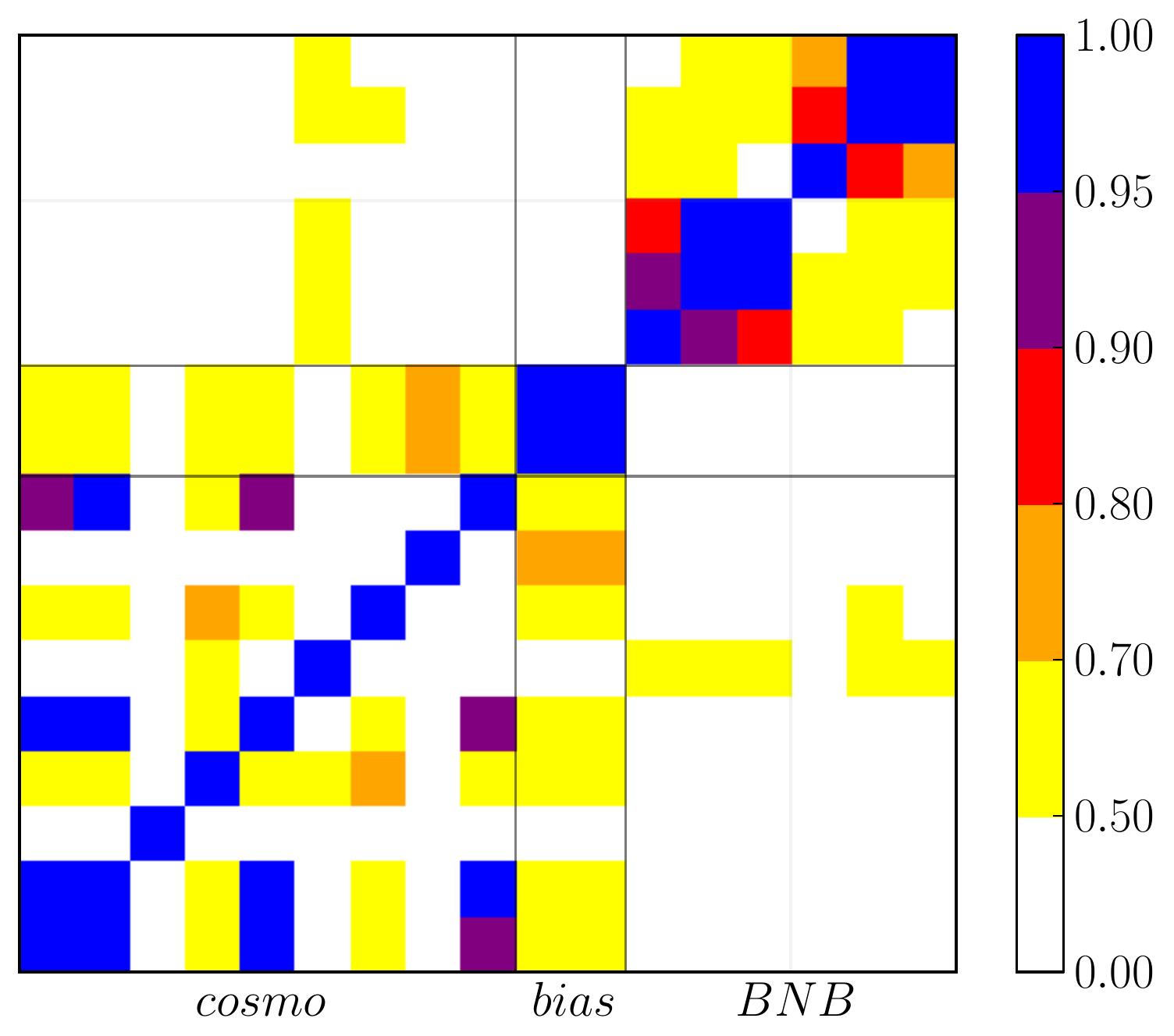}
	\includegraphics[width=\columnwidth]{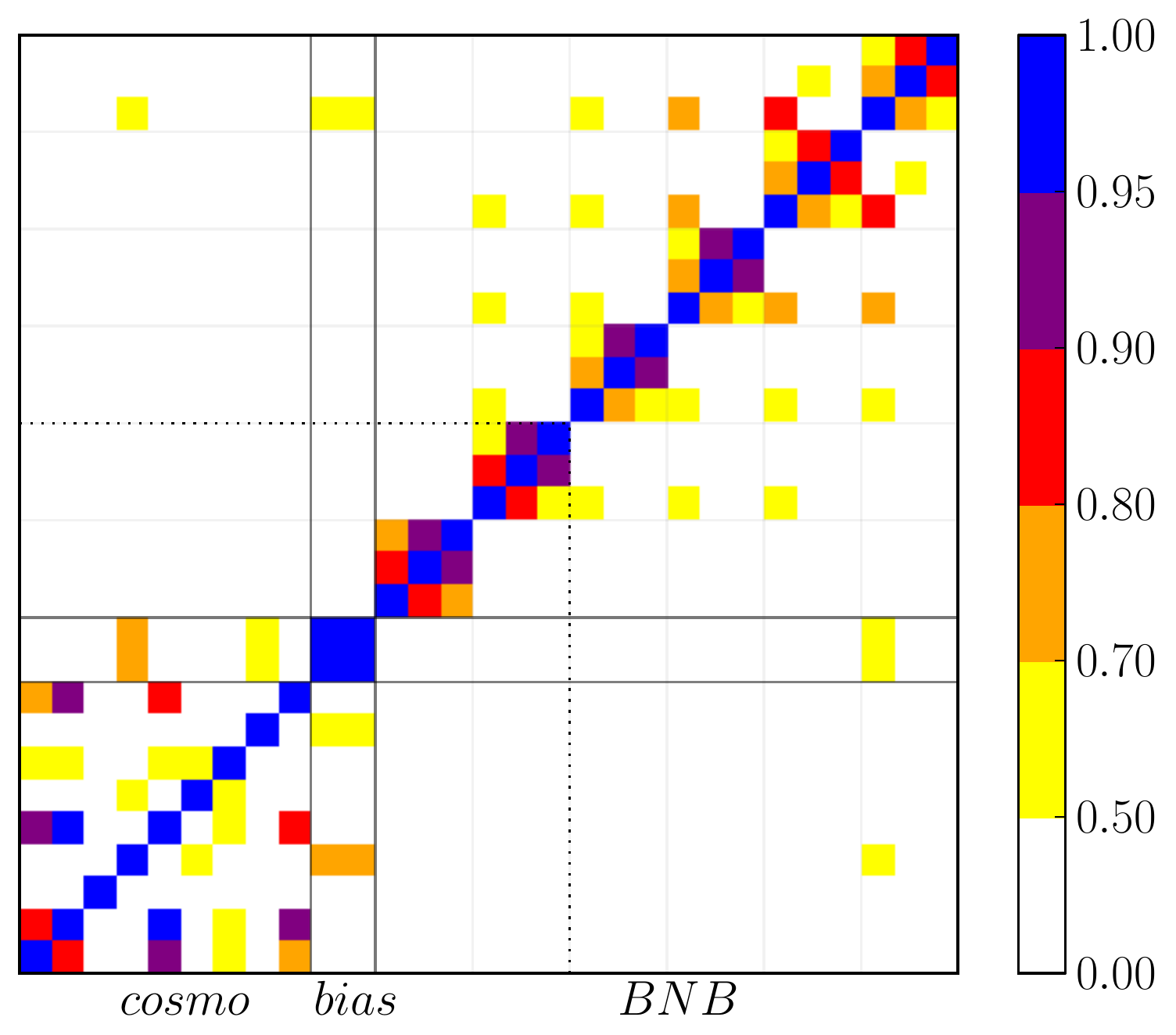}\\ 
	\includegraphics[width=\columnwidth]{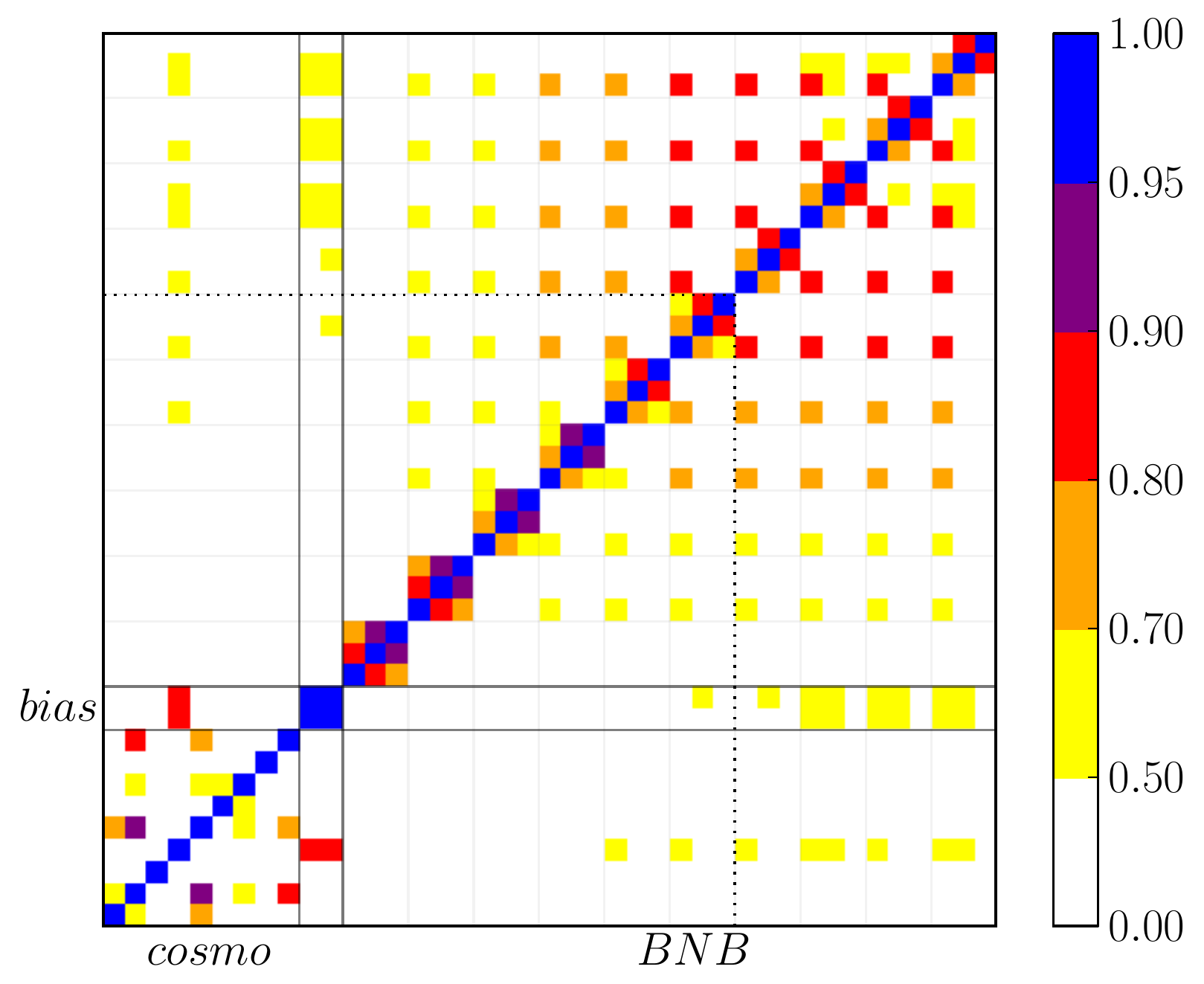}
	\includegraphics[width=\columnwidth]{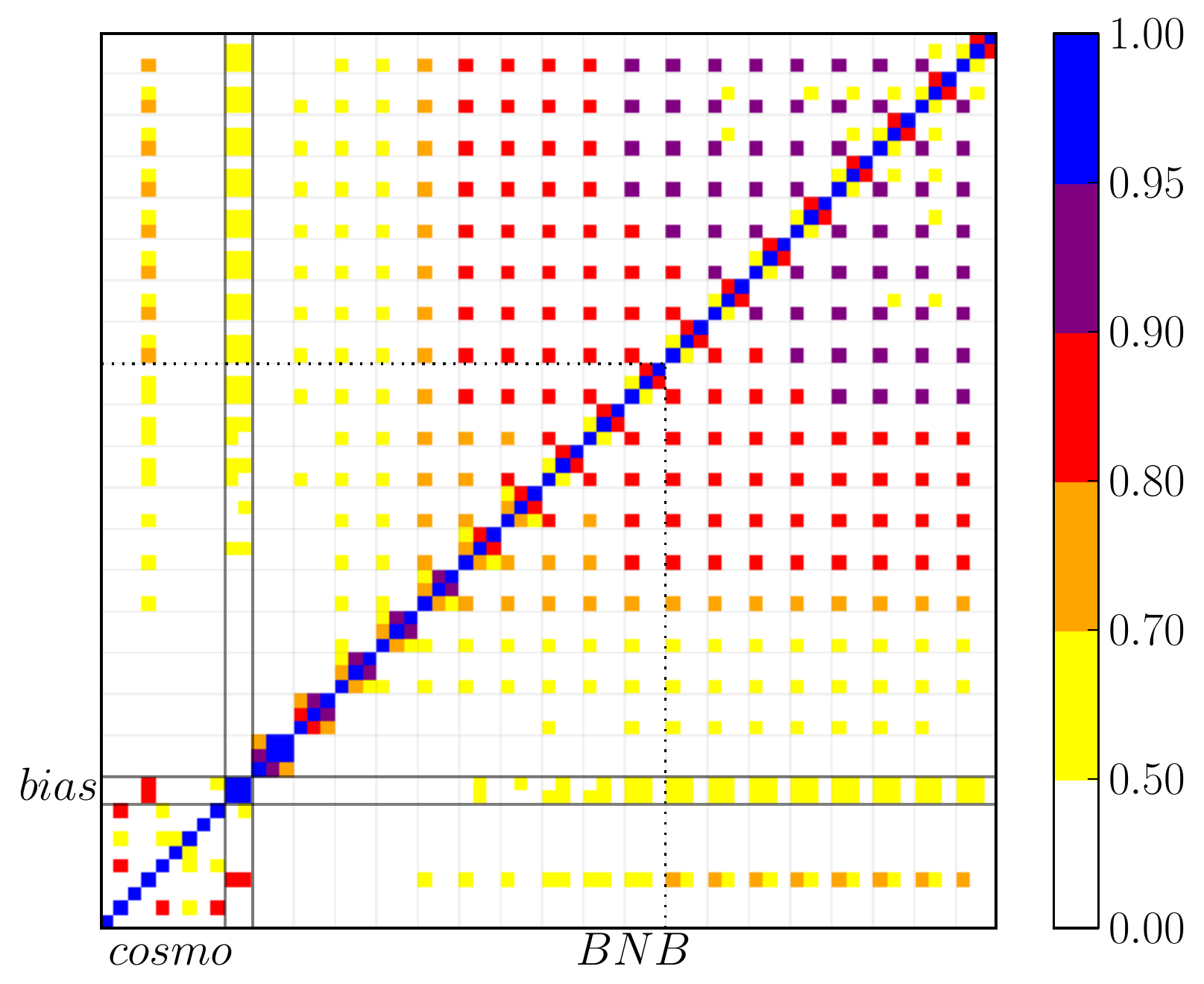}\\ 
	\includegraphics[width=\columnwidth]{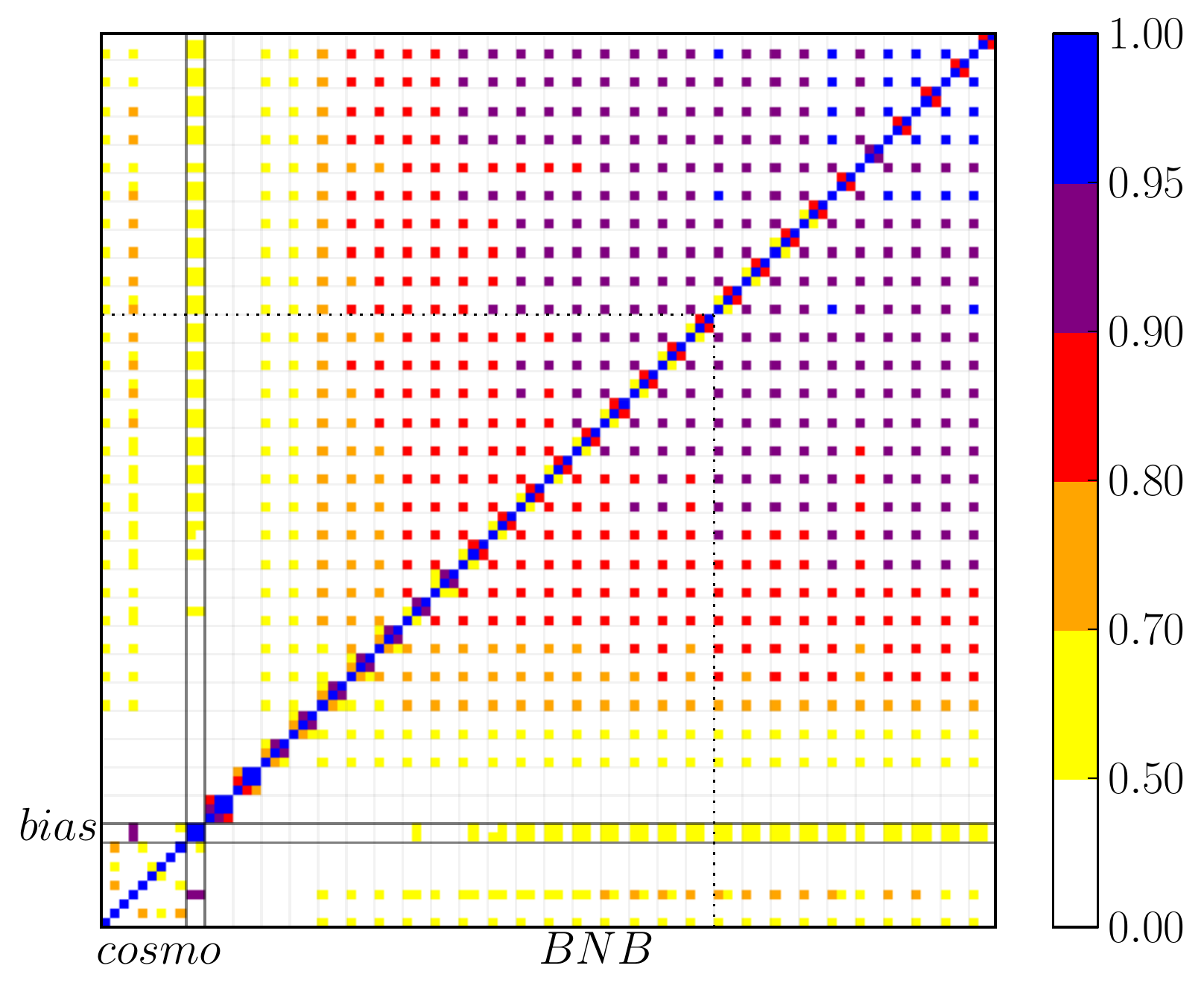}
	\includegraphics[width=\columnwidth]{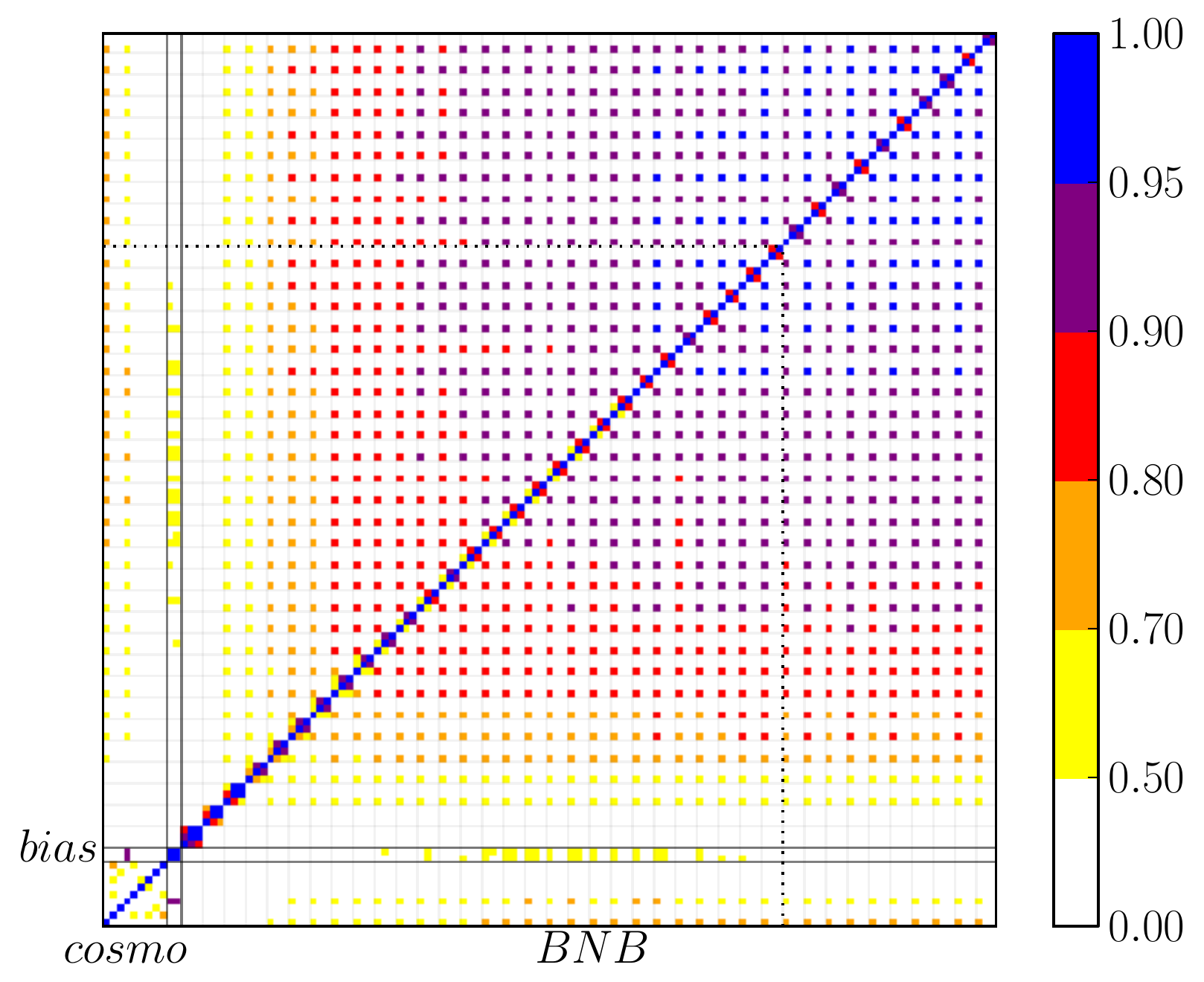}
	\caption{The absolute value of the correlation coefficients 
are shown for $k_{\rm max} = 0.1$, 0.2 (top row), 0.3, 0.5 (middle), 
0.75, $1\,h$/Mpc (bottom row). White space indicates that the correlation 
coefficient $|r_{ij}|<0.5$. The blocks of cosmology, bias, and 
baryonic/nonlinear/scale dependent bias (BNB) parameters are labeled. 
Dotted lines indicate the limit of the 
$k$ bins from the previous panel in the series. 
} 
	\label{fig:rcorr}
\end{figure*}

The final parameter constraints are convolutions of all the Fisher derivatives
and their interplay.  To illustrate the role of higher $k$ modes in breaking
covariances, we plot the absolute value of the correlation matrix, $|r_{ij}| =
|C_{ij}|/\sqrt{C_{ii}C_{jj}}$, where $C=F^{-1}$ is the parameter covariance
matrix. We focus on the high correlation coefficients (note diagonal entries
are 1 by definition). Figure~\ref{fig:rcorr} shows these for various
$\kmax$. For clarity the matrix is divided into blocks, with the lower left
containing the cosmological parameters, the next (small) block the fiducial
bias parameters $\{ b^0_{\rm ELG}, b^0_{\rm LRG} \}$, and the upper right
block the BNB parameters $\{B_k, c_{1,k}, c_{2,k}\}$.  The offdiagonal blocks
contain the cross terms.

Note that the cosmology parameters become progressively less correlated 
with each panel at higher $\kmax$, with the cosmology block becoming both 
sparser and lighter colored. The BNB parameters, however, 
retain their correlation (the dotted lines show the size of the matrix 
from the previous $\kmax$ step, making it clearer to compare the $k$ bins). 
Thus, going to higher $\kmax$ delivers two significant effects in favor of 
improving cosmology constraints -- mode statistics and degeneracy breaking 
-- while the model independent binning approach removes the worry of 
misestimating the nonlinear or baryonic behavior. 

\begin{figure}[htbp!]
\includegraphics[width=\columnwidth]{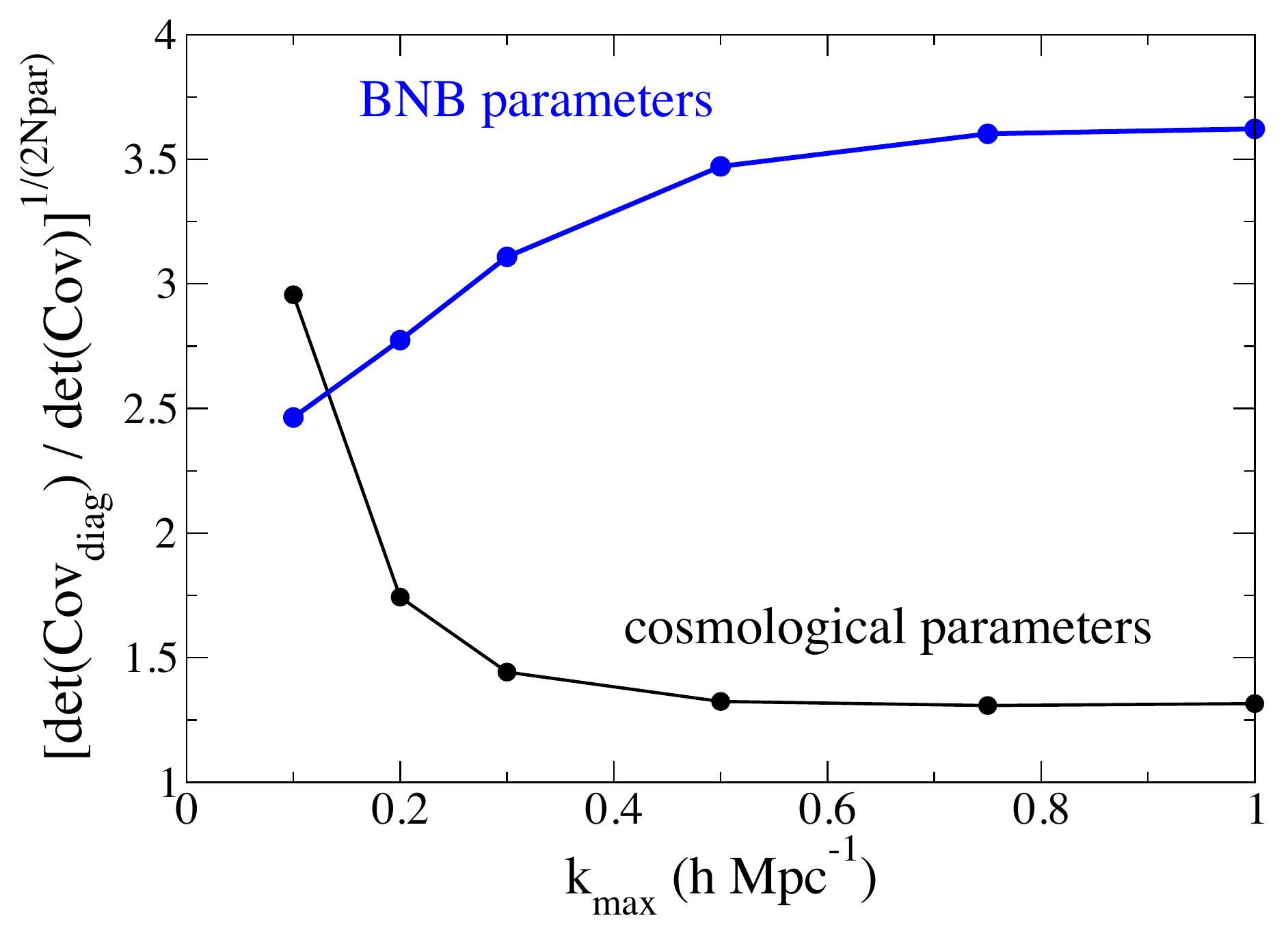} 
\caption{The decreasing correlation among cosmological parameters as 
information is included from beyond the linear regime, to higher $\kmax$, 
is demonstrated by taking the volume of their $N$-dimensional ellipsoid 
including covariances relative to the volume defined by only the product 
of their variances (see Eq.~\ref{eq:vsub}). Baryonic/nonlinear/scale 
dependent bias parameters, however, are significantly correlated. 
} 
\label{fig:detc} 
\end{figure}

A more compact illustration of the reduction in covariance among cosmological 
parameters with increasing $\kmax$ appears in Fig.~\ref{fig:detc}. Here we 
consider the volumes of the parameter-space ellipsoids in two subspaces: the
9-parameter cosmology space, and the BNB parameter space. In each case, 
we compare how much the volume (square root of the determinant of the covariance matrix)
increases in the hypothetical case that the parameters are completely 
uncorrelated, versus the actual correlated case. This is a generalized measure of how much correlation there is
in the subspace. Because this ratio strongly increases with increasing dimensionality
of the subspace, $N_{\rm subspace}$, we also take the $N_{\rm subspace}$-th root of the ratio, making
it a ``one-parameter equivalent'' increase in volume.
The ratio is therefore defined as
\be 
R_{\rm corr}\equiv
\left(\frac{\det\,({\rm diag}\, C_{\rm subspace})}{\det\, (C_{\rm subspace})}\right)^{1/(2N_{\rm subspace})}. 
\label{eq:vsub} 
\ee 
When $R_{\rm corr}$ is near unity, the parameter subspace is 
substantially decorrelated; when it is much larger than unity then 
covariances play an important role. We see from Fig.~\ref{fig:detc} that 
indeed the cosmological parameters become increasingly uncorrelated as 
$\kmax$ increases, potentially allowing rapid improvement in their 
constraints.

Finally, we must assess whether the numerous extra fit parameters for the 
power spectrum at high $k$ degrade the cosmology estimation. By looking at 
the cross terms in the cosmo-BNB bands, we see that there is little 
covariance (the main, though mild, correlation is with $\Omega_K$). 
Thus we expect that the cosmology estimation precision should improve 
significantly by using these higher $k$ modes, within this marginalized 
bin approach. 

\begin{figure}[htbp!]
\includegraphics[width=\columnwidth]{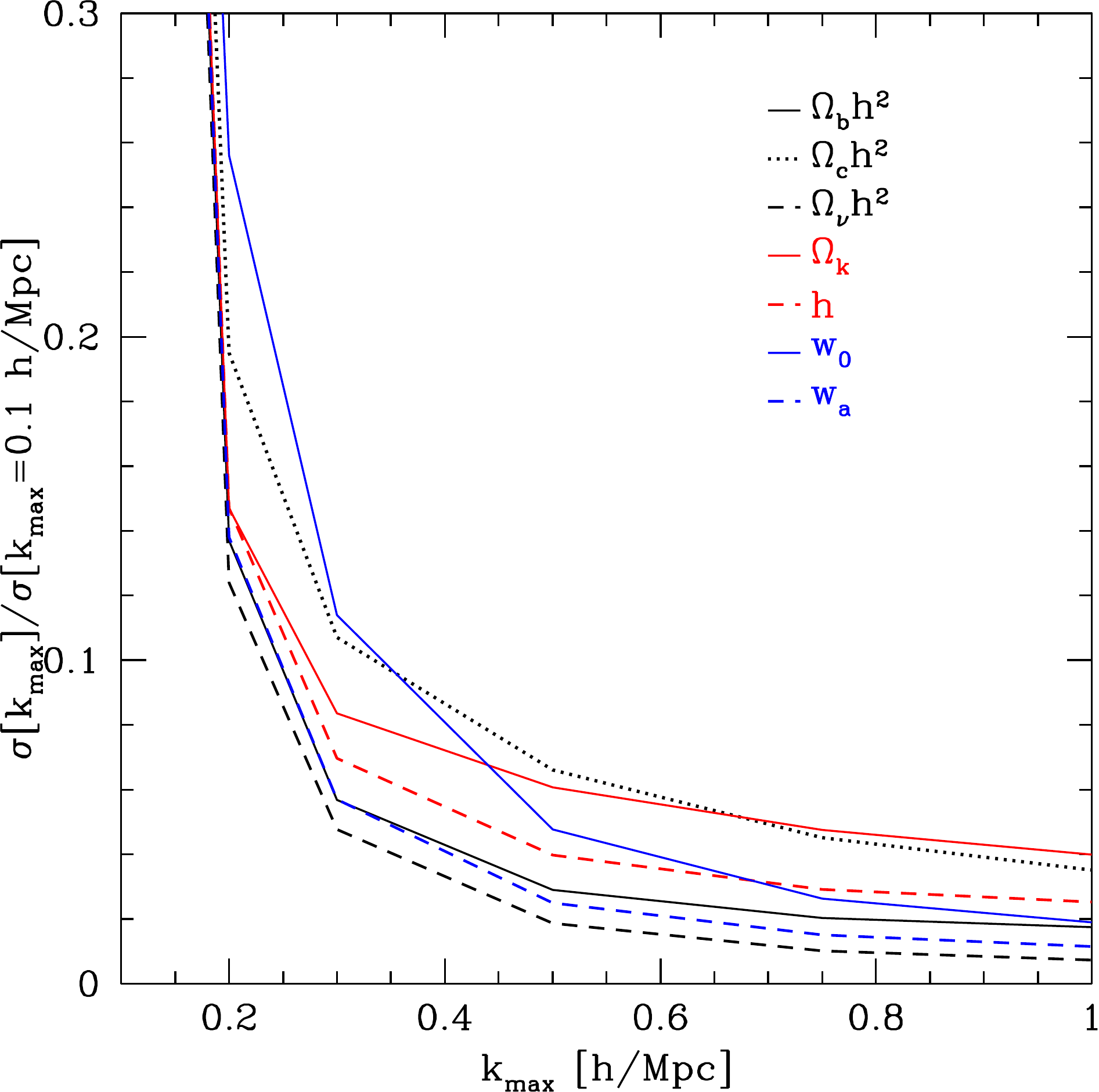} 
\caption{The ratio of the parameter uncertainty when using information out 
to $\kmax$, relative to using $\kmax=0.1\,h$/Mpc (so all curves go to 1 at 
$\kmax=0.1\,h$/Mpc), is plotted vs $\kmax$ for 
several cosmology parameters. The constraints 
rapidly improve, despite the extra parameters in each $k$ bin for the 
baryonic/nonlinear/scale dependent bias effects. 
} 
\label{fig:ratio} 
\end{figure}

Figure~\ref{fig:ratio} demonstrates this result. The cosmology parameter
estimation improves dramatically when going beyond the linear regime, despite
the addition of 30, 54, and 114 extra parameters for $\kmax=0.3$, 0.5,
$1\,h$/Mpc. As expected, due to the lingering covariances, $\Omega_K$ is the
parameter that improves most slowly at higher $k$, while the clear scale
dependence of neutrino mass means that it improves most rapidly.
Table~\ref{tab:sigmas} summarizes the results (again, this should be
interpreted predominantly in terms of information content, not actual
constraints since the real space power spectrum is not truly an observable
quantity in a survey).

\begin{table*}[!t]
\ra{1.3}
\begin{tabular*}{0.9\textwidth}{@{\extracolsep{\fill} }l | c c c c c c c c c c c} 
$\kmax$ & $10^3\Omega_b h^2$ & $10^3\Omega_{\rm CDM}h^2$  & $10^4\Omega_\nu h^2$ & $\Omega_K$ & $100\,h$ & $w_0 $ & $w_a$ & $10^9 A_s$ & $n_s$ & $b^0_{\rm ELG} $  & $b^0_{\rm LRG}$ \\ \hline
0.1 & 4.0 & 16.0 & 4.2 & 0.21 & 4.8 & 0.23 & 1.14 & 0.53 & 0.063 & 0.17 & 0.33 \\ 
0.2 & 0.55 & 3.2 & 0.53 & 0.030 & 0.70 & 0.059 & 0.16 & 0.082 & 0.019 & 0.023 & 0.045 \\ 
0.3 & 0.23 & 1.8 & 0.20 & 0.017 & 0.33 & 0.026 & 0.065 & 0.037 & 0.013 & 0.013 & 0.026 \\ 
0.5 & 0.12 & 1.1 & 0.079 & 0.013 & 0.19 & 0.011 & 0.028 & 0.016 & 0.0097 & 0.0094 & 0.020 \\ 
0.75 & 0.082 & 0.74 & 0.043 & 0.0098 & 0.14 & 0.0060 & 0.017 & 0.0095 & 0.0075 & 0.0075 & 0.016 \\ 
1.0 & 0.070 & 0.58 & 0.031 & 0.0082 & 0.12 & 0.0044 & 0.013 & 0.0071 & 0.0063 & 0.0063 & 0.013 
	  \end{tabular*}
  \caption{The $1\sigma$ cosmology and fiducial bias parameter uncertainties 
are given for the baseline survey using information out to various
$\kmax$. 
} 
 \label{tab:sigmas}
\end{table*}

Besides the cosmological parameters, the galaxy linear bias parameters 
become well determined for $\kmax\gtrsim0.2\,h$/Mpc, reaching the $\sim1\%$ 
level. Interestingly, the $k$-bin parameters -- representing the effects of 
baryons, nonlinearity, and (scale dependent) galaxy bias on the galaxy 
power spectrum, not already captured by the Halofit and linear bias models --
self calibrate to a large degree. The fractional uncertainties on the 
bin amplitudes $B_k$ tend to be 0.8--1.6\%, while the redshift-dependence 
parameters $c_{1,k}$ and $c_{2,k}$ are determined to about 0.01--0.02, for $\kmax\gtrsim0.2\,h$/Mpc.

\section{Testing Alternatives} \label{sec:fids} 

The approach of fitting for binned deviations from the Halofit power 
spectrum gives successful results. Arbitrary deviations, however, could 
mock up a change in any cosmological parameter, so we should test that 
the constraints we imposed -- on the redshift dependence of the nonlinear 
deviations and assuming small deviations from Halofit (so the Fisher 
analysis is in its region of validity) -- yield reasonably generic results. 

We therefore investigate the effect of altering our baseline approach in 
three different ways: loosening the redshift dependence, adopting a 
different fiducial scale dependence, and allowing a mixing between the 
scale and redshift dependence.

\subsection{Redshift dependence} \label{sec:fidz} 

Although a second order polynomial in redshift seems a reasonable 
treatment for the influence of BNB effects on the power spectrum 
at $z<2$, we here test its influence by extending the freedom further 
with a cubic term. That is, Eq.~(\ref{eqn:Mkz}) now becomes 
\be 
M(k,z)=(1+c_{1,k}z+c_{2,k}z^2+c_{3,k}z^3)\,B_k \ . 
\ee 
This adds one parameter per $k$ bin, giving a total of 83 fit parameters 
for $\kmax=0.5\,h$/Mpc say. The fiducial remains $B_k=1$, $c_{i,k}=0$. 

\begin{figure}[htbp!]
\includegraphics[width=\columnwidth]{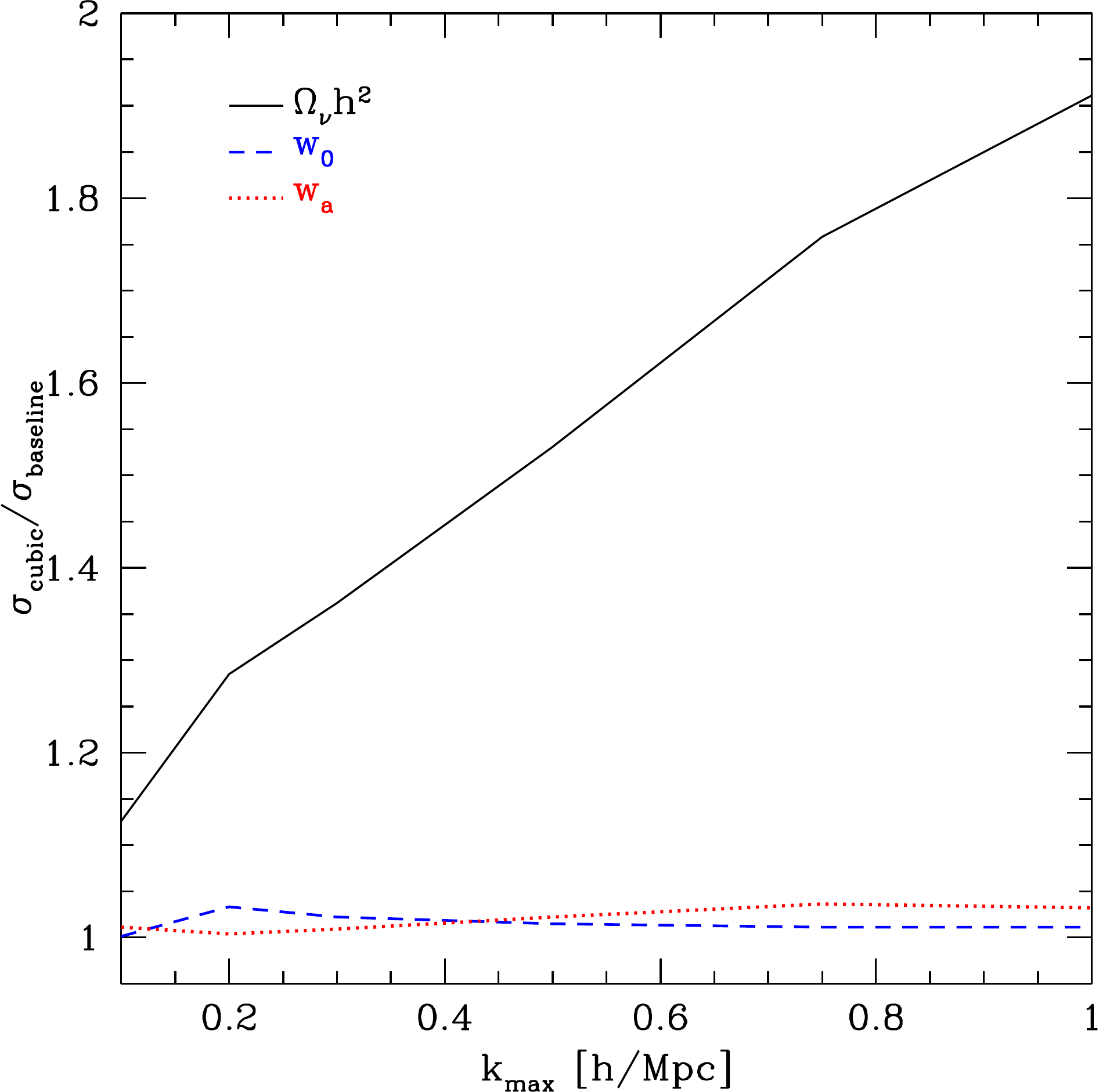} 
\caption{The ratio of the parameter uncertainty when allowing for an extra, 
cubic redshift fit parameter $c_{3,k}$ in the scale dependence, relative to 
our baseline fiducial, is plotted vs $\kmax$ for several cosmology 
parameters. Only the neutrino energy density shows a significant effect. 
} 
\label{fig:fidz} 
\end{figure}

We find that most of the cosmology results are affected little, with less 
than 10\% change in the parameter estimation uncertainties. The main 
exception is the neutrino energy density constraint, as seen in 
Fig.~\ref{fig:fidz}. Since the neutrino free streaming scale is dependent 
on both scale and redshift, the extra redshift dependence in the scale 
dependent BNB factor allows greater covariance, weakening the 
constraint by less than a factor of 2. The BNB parameters 
themselves are also less well determined, by factors of up to 1.7 for 
$B_k$ (so uncertainties of 1.3--2\%); uncertainties on the three $c_{i,k}$ 
become up to $\sim$0.07, 0.09, 0.03.

\subsection{Scale dependence} 

To treat the BNB effects in as model 
independent a manner as possible, we allowed free floating bins in $k$ 
to describe deviations from the (dark matter plus neutrino plus linear bias) 
Halofit 
prescription. Since Fisher analysis is only accurate for small deviations 
around the fiducial, then we may not have accurately captured the effect 
of large baryonic deviations. Therefore we now adopt a fiducial power 
spectrum that attempts to include the baryonic effects to high $k$ using 
the recent work of \cite{Mohammed:2014lja}. This adds corrections as a 
polynomial in $k$, whose form is motivated by the Taylor expansion of 
the 1-halo term in wavenumber $k$. 
Specifically, rather than take a fiducial of $B_k=1$ 
in each bin, we here use 
\be 
B_k=1+(1.33+5.96\,k^2-4.63\,k^4)\,G(k) \ , 
\ee 
based on the simulation results of \cite{Mohammed:2014lja}, where in this 
formula $k$ is in dimensions of $h$/Mpc, 
$G(k)=F(k)/P_{\rm no\,bary}(k)$, and $F(k)$ is given by their Eq.~(30). 

We find that this change in fiducial has minimal effect on our results. 
The maximum fractional change in a parameter uncertainty is 2.7\%, with 
most alterations below 1\%. Thus, our results appear robust to this 
modification.

\subsection{Mixing redshift and scale dependence} 

Some physical effects on the power spectrum may not be well treated by
multiplicative factors of scale dependence times redshift dependence
(e.g.\ redshift dependent physical scales such as from baryonic feedback or
neutrino free streaming).  While our BNB approach does allow mixing of scale
and redshift dependence, the fiducial values $c_{i,k}=0$ mean this only enters
through the Fisher derivatives, because the fiducial $M_{\rm fid}(k,z)$ in
Eq.~(\ref{eqn:Mkz}) then becomes independent of $z$.  As a simplistic
exploration of the possible impact of such an effect, we adopt a fiducial
model that makes the redshift dependence vary with scale. Specifically,
\be 
c_{i,k}=c_{i,\infty}\,\left(1-e^{-k/k_\star}\right) \ , 
\ee 
and rather than the baseline fiducial $c_{i,k}=0$ we set $c_{1,\infty}=2$, 
$c_{2,\infty}=1$, $k_\star=0.3\,h$/Mpc. Thus the fiducial redshift dependence 
of the BNB effects on the power spectrum stays at the baseline at low 
$k$, but transitions to $(1+z)^2$ at high $k$. 
This is intended purely as a toy model, with the characteristic that the 
fiducial deviations are stronger at higher redshift, to test the baseline 
Fisher analysis. 

We find that this mixed fiducial increases correlations among the BNB 
parameters but has no deleterious effect on estimation of cosmological 
parameters. Indeed, since for $k\gtrsim k_\star$ the fiducial power 
spectrum is boosted by a factor $(1+z)^2$, this enhances $nP_{\rm fid}$ 
at high $k$ and reduces shot noise, improving most cosmology parameter 
estimates by $\sim25\%$.

\section{Conclusions} \label{sec:concl} 

Large scale structure surveys provide a rich array of information on 
cosmology and astrophysics on scales from the survey size down to small 
scales, or high wavenumbers. Using this data beyond the linear regime, 
where we do not fully comprehend all the important physics, is a challenge 
but one with rich rewards for cosmological understanding. Here 
we have analyzed how uncertainties in the theoretical prediction for the 
galaxy power spectrum out to a maximum wavenumber $k_{\rm max}$ impact 
cosmological parameter constraints, and how we can mitigate the uncertainties 
without biasing the results. 

We included three physical effects: baryonic modifications, 
nonlinearities, and scale-dependent bias -- referred to jointly as the
BNB effects.  To guard against bias from improperly assuming specific 
functional forms, we employed a very flexible, nearly model-independent 
description of the BNB effects that allows scale- and redshift-dependence, 
described by between 6 and 114 additional fit parameters, depending on 
$k_{\rm max}$. 

Despite the addition of these BNB parameters we could still obtain 
excellent constraints on cosmology. In fact, the cosmological constraints 
improve rapidly with
increasing $k_{\rm max}$, despite the growing number of extra
astrophysical parameters to marginalize over. We traced this improvement 
to two mutually
reinforcing effects: a well-known fact that the information content increases
sharply with $k_{\rm max}$ due to more modes, but also the key property 
that the cosmological parameter correlations 
weaken as smaller-scale information provides leverage to break degeneracies. 
We tested this conclusion against different assumptions for the
form of the BNB sector, altering the fiducial redshift-, scale-, and mixing 
of redshift-scale dependence, and 
found it to be quite robust. 

These results agrees qualitatively with other, previous work which shows that
cosmological data, and particularly the two-point correlation function, can be 
remarkably robust with respect to self-calibrating nuisance parameters. In
other words, one can add a number of judiciously chosen nuisance parameters  
that describe the systematic uncertainties, and these parameters, together with
cosmological ones, can be internally determined from the data. For example,
\cite{Wu:2013pia} showed that the three-dimensional galaxy clustering can 
be used to self-calibrate the 
parameters describing how galaxies occupy halos (the Halo Occupation
Distribution), leading to improvements on small scales despite the rather
aggressive modeling of the systematics.  

Measuring a large number of nuisance parameters will likely not be
feasible due to practical considerations -- even if one were able to constrain
of order 100 nuisance parameters, doing so might not be robust or convenient
in the presence of purely observational and instrument-related systematics
which require special care and computational resources in their own right.
Fortunately, simpler approaches may bear fruit in the near future. For example,
a careful investigation of the effects of baryonic systematics based on a
suite of numerical simulations seems to indicate that the systematics span a
subspace in the observable (say, the weak lensing angular power spectrum)
that is rather orthogonal to that spanned by the cosmological parameters 
\cite{Eifler:2014iva}. Therefore, provided one has successfully modeled the
systematics, one can reasonably expect to self-calibrate them or marginalize
over their parameters and still be able to constrain the cosmology with
excellent accuracy. Again a key issue is guarding against biased results by 
enhancing both the model independence and the flexibility of the treatment 
of the BNB effects. 

The information content of cosmological data well beyond the linear scale is 
high. This provides strong motivation to push to large $\kmax$ while dealing 
robustly with the baryonic/nonlinearity/scale dependent bias effects 
masking this signal. The substantially model independent, marginalization 
approach we present could be a harbinger of rich rewards in cosmological 
knowledge, without problematic biasing of results, from robust analysis of 
next generation large-scale-structure measurements.

\acknowledgments 

JB thanks LBNL for hospitality during part of this work. His work has 
been supported in part by the DOE grant DE-SC0010386 at Dartmouth. 
DH is supported by the DOE grant under contract DE-FG02-95ER40899 and 
NSF under contract AST-0807564. 
EL is supported in part by DOE grant DE-SC-0007867 and DE-AC02-05CH11231, 
and NASA. 

\appendix 

\section{Binning robustness test} \label{sec:apx} 

The approach taken to treat the BNB uncertainty in the power spectrum is to 
allow model independent, free floating bin parameters in $k$, with redshift 
dependence given by a second order polynomial with free coefficients. In 
Sec.~\ref{sec:fidz} we tested the impact of allowing further redshift 
dependence. We would now also like to 
test the influence of the binning. We chose our 
fiducial value of $\Delta k=0.025\,h$/Mpc to avoid interaction with the baryon 
acoustic harmonic of $k\approx0.06\,h$/Mpc. Here we examine the effect 
on the cosmological parameter estimation by using $\Delta k=0.01\,h$/Mpc. For 
$\kmax=0.5\,h$/Mpc this implies 146 total fit parameters. 

\begin{figure}[htbp!]
\includegraphics[width=\columnwidth]{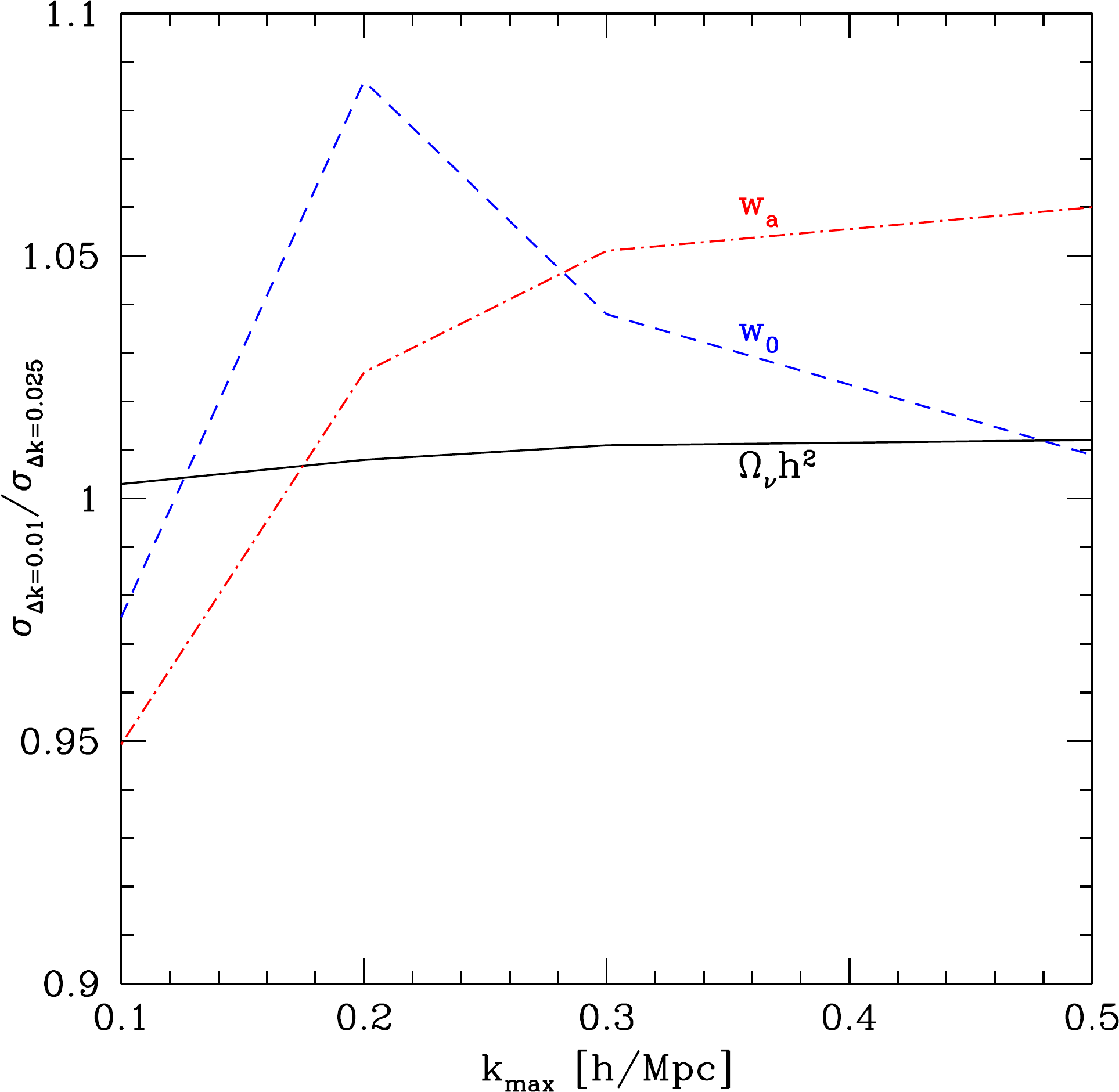} 
\caption{The ratio of the parameter uncertainty when allowing for a finer 
binning, $\Delta k=0.01\,h$/Mpc, relative to 
our baseline fiducial $\Delta k=0.025\,h$/Mpc, is plotted vs $\kmax$ for several cosmology 
parameters. The cosmology estimation remains robust. 
} 
\label{fig:sigdk} 
\end{figure}

We find that the cosmological parameter uncertainty estimation remains 
quite robust. The greatest change is an increase in the $n_s$ uncertainty by 
50\% at $\kmax=0.3\,h$/Mpc (addition of CMB data would constrain 
$n_s$, decreasing the dependence of its estimation on $k_{\rm max}$). 
For the BNB parameters, constraints weaken due to increased 
covariance. However our main goals are the estimation of dark energy and 
neutrino parameters.  Figure~\ref{fig:sigdk} shows that these 
uncertainties change by less than 10\% with the change in binning, so the 
results we have presented appear robust.

\bibliography{refs}

\end{document}